%% file: templatePRIME.tex
\documentclass{article}

\usepackage{PRIMEarxiv}

\usepackage[utf8]{inputenc} 
\usepackage[T1]{fontenc}    
\usepackage{hyperref}       
\usepackage{url}            
\usepackage{booktabs}       
\usepackage{amsfonts}       
\usepackage{nicefrac}       
\usepackage{microtype}      
\usepackage{lipsum}
\usepackage{fancyhdr}       
\usepackage{graphicx}       
\graphicspath{{media/}}     

\title{PILLAR: an AI-Powered Privacy Threat Modeling Tool
}

\author{
	Majid Mollaeefar $^a$, Andrea Bissoli $^a$, and Silvio Ranise $^{a,}$ $^b$ \\
	$^a$ FBK-Center for Cybersecurity, Trento, Italy.\\
	$^b$ Department of Mathematics, University of Trento, Trento, Italy.\\
	\\
	\texttt{\{mmollaeefar, abissoli, ranise\}@fbk.eu} \\
}

\begin{document}
	\maketitle

	\begin{abstract}
		The rapid evolution of Large Language Models (LLMs) has unlocked new possibilities for applying artificial intelligence across a wide range of fields, including privacy engineering. As modern applications increasingly handle sensitive user data, safeguarding privacy has become more critical than ever. To protect privacy effectively, potential threats need to be identified and addressed early in the system development process. Frameworks like LINDDUN offer structured approaches for uncovering these risks, but despite their value, they often demand substantial manual effort, expert input, and detailed system knowledge. This makes the process time-consuming and prone to errors.
		Current privacy threat modeling methods, such as LINDDUN, typically rely on creating and analyzing complex data flow diagrams (DFDs) and system descriptions to pinpoint potential privacy issues. While these approaches are thorough, they can be cumbersome, relying heavily on the precision of the data provided by users. Moreover, they often generate a long list of threats without clear guidance on how to prioritize them, leaving developers unsure of where to focus their efforts.
		In response to these challenges, we introduce PILLAR (Privacy risk Identification with LINDDUN and LLM Analysis Report), a new tool that integrates LLMs with the LINDDUN framework to streamline and enhance privacy threat modeling. PILLAR automates key parts of the LINDDUN process, such as generating DFDs, classifying threats, and prioritizing risks. By leveraging the capabilities of LLMs, PILLAR can take natural language descriptions of systems and transform them into comprehensive threat models with minimal input from users, reducing the workload on developers and privacy experts while improving the efficiency and accuracy of the process.
	\end{abstract}

	\keywords{Privacy Threat Modeling \and Large Language Model \and Risk Assessment }

	\input{Sections/1-Introduction}
	\input{Sections/2-Background}
	\input{Sections/3-Tool}

	\input{Sections/5-Relatedwork}

	\input{Sections/6-Conclusion}

	\bibliographystyle{unsrt}  
	\bibliography{references}

\end{document}

%% file: Sections/1-Introduction.tex
\section{Introduction}
\label{sec:intro}
In today's digital landscape, privacy has become a paramount concern as applications increasingly handle sensitive user data. Ensuring robust privacy protection requires identifying and mitigating potential privacy threats during the early stages of system development. Privacy threat modeling frameworks, such as LINDDUN~\cite{deng2011privacy}, provide structured approaches to uncover and address these threats. However, despite their structured methodologies, these frameworks often require significant manual effort, expert knowledge, and detailed system descriptions, which can be time-consuming and prone to oversight.

Existing privacy threat modeling approaches, including LINDDUN, involve creating and analyzing detailed data flow diagrams (DFDs) and other system descriptions to identify potential privacy risks. While effective, this process is often cumbersome and depends heavily on the accuracy and completeness of the input provided by users. Additionally, these methods can struggle with the prioritization of identified threats, leading to an overwhelming list of potential issues without clear guidance on which to address first.

To address these challenges, we propose \textbf{PILLAR} (Privacy risk Identification with the LINDDUN and LLM Analysis Report), a novel tool that integrates Large Language Models (LLMs) with the LINDDUN framework to automate and enhance the privacy threat modeling process. PILLAR is designed to simplify the identification and analysis of privacy threats by automating key aspects of the LINDDUN methodology, such as the generation of DFDs, the categorization of threats, and the prioritization of risks. By leveraging LLMs, PILLAR can process natural language descriptions of systems to produce detailed threat models with minimal user input, reducing the burden on developers and security researchers.

Our tool introduces an innovative approach to threat modeling by simulating multi-agent collaboration among virtual experts, each focused on different aspects of privacy threats. 
In PILLAR, virtual agents, such as a privacy expert or a developer, communicate and debate privacy risks using distinct but complementary perspectives.
This structure reflects the \textit{cooperative communication} paradigm outlined by Guo et al. (2024) \cite{guo2024large}, 
where multiple agents cooperate to achieve a shared goal through interactions and knowledge sharing. By simulating multiple rounds of communication between agents, PILLAR reduces the likelihood of overlooking critical privacy risks, mirroring the real-world collaborative efforts between stakeholders in privacy threat modeling. This approach enables the tool to capture a broader range of potential threats by allowing agents to deliberate and adjust their insights iteratively, leading to more thorough risk assessments.


In this work, we present the design and implementation of PILLAR, which introduces several key innovations to enhance and streamline the privacy threat modeling process:

\begin{itemize}
	\item \textbf{Automated Application Description}: PILLAR enables users to input a natural language description of their application, including details on the type (e.g., mobile app, web app), the data it collects, and its data policies. This automation helps reduce manual effort and ensures accurate system descriptions.
	
	\item \textbf{Data Flow Diagram (DFD) Management}: Users can upload, edit, generate from an image or description, and download DFDs, which are essential for identifying privacy threats. This flexibility streamlines the modeling and visualization of system data flows.
	
	\item \textbf{LINDDUN Analysis Automation}: PILLAR facilitates quick and structured LINDDUN analyses within the platform, simplifying the process of identifying potential privacy risks.
	
	\item \textbf{LINDDUN Go Simulation}: By incorporating LLM agents, PILLAR offers LINDDUN Go simulations to evaluate privacy threats dynamically, allowing for scenario-based analysis with multiple virtual agents working collaboratively.
	
	\item \textbf{LINDDUN Pro Methodology}: For more advanced threat identification, PILLAR integrates the LINDDUN Pro methodology, allowing users to perform precise and thorough threat analyses using DFDs.
	
	\item \textbf{Impact Assessment and Control Measures}: PILLAR provides tools for assessing the impact of identified threats and suggests control measures based on established privacy patterns, guiding users in mitigating risks effectively.
	
	\item \textbf{Comprehensive Report Generation}: Finally, PILLAR allows users to generate and download a detailed report summarizing the entire privacy threat modeling process, including identified threats, impact assessments, and recommended control measures, aiding in documentation and compliance efforts.
\end{itemize}

These contributions demonstrate PILLAR’s potential to automate, simplify, and enhance the privacy threat modeling process, making it more accessible and efficient for developers, privacy experts, and security researchers alike. 
Furthermore, we implemented PILLAR as a web-based tool which offers a friendly user interface (UI) to its users by utilizing an interactive web application framework. 
 
The remainder of this paper is organized as follows: Section~\ref{sec:back} provides a necessary background on threat modeling and Large Language Models. Section~\ref{sec:tool} details the architecture and functionality of the PILLAR tool. Section 4 presents our case studies and evaluation results. Related work is discussed in Section~\ref{sec:related}. Finally, Section~\ref{sec:conclusion} concludes with a discussion of future work and potential enhancements.


%

%% file: Sections/2-Background.tex
\section{Background}
\label{sec:back}
In this section, we focus on two primary areas: the importance of threat modeling in cybersecurity and privacy, and the application of Large Language Models to enhance the threat modeling process. By combining these methodologies, we aim to address the challenges in identifying and mitigating privacy risks efficiently and effectively.

\subsection{Threat Modeling}
Threat modeling is a foundational practice in cybersecurity and privacy management, providing a structured methodology to identify, assess, and mitigate potential risks within a system, and support in ensuring compliance with regulations such as the General Data Protection Regulation~\cite{regulation2016} (GDPR). It is crucial for several reasons: 
First, it enables organizations to identify vulnerabilities by analyzing system architecture and potential attack vectors, helping to pinpoint weak spots that could be exploited by adversaries. Second, threat modeling allows for the prioritization of risks based on both their potential impact and likelihood, ensuring that resources are allocated effectively to address the most critical vulnerabilities. 
Third, it fosters clear communication among various stakeholders, including developers, security teams, and management, ensuring that all parties understand the risks and the corresponding mitigation strategies.
Moreover, threat modeling supports compliance with regulatory requirements, as many legal frameworks mandate comprehensive risk assessments. 
Finally, threat modeling is not a one-time effort, it is a continuous process that should be integrated into the software development life cycle (SDLC), allowing for ongoing assessment and adaptation to emerging threats.

There are several established methodologies for conducting threat modeling, each offering a distinct approach to security and privacy risk identification. These methodologies are widely adopted in both industry and academia:

\begin{itemize}
    \item \textbf{STRIDE}~\cite{shostack2014}: This framework categorizes security threats into six distinct types: Spoofing, Tampering, Repudiation, Information Disclosure, Denial of Service (DoS), and Elevation of Privilege. STRIDE is predominantly used to identify security vulnerabilities in software applications.
    
    \item \textbf{PASTA}~\cite{ucedavelez2015risk}: The Process for Attack Simulation and Threat Analysis (PASTA) emphasizes simulating potential attacks to understand their impact on business objectives. This methodology is more business-driven, focusing on how security threats affect an organization’s goals and operations.
    
    \item \textbf{LINDDUN}~\cite{wuyts2015linddun}: Specifically designed for privacy threat modeling, LINDDUN identifies privacy-specific risks, including Linkability, Identifiability, Non-repudiation, Detectability, Disclosure of Information, Unawareness, and Non-compliance. It offers a structured process for assessing privacy risks in systems, aligning with regulations like the General Data Protection Regulation (GDPR).
\end{itemize}

\subsection{The LINDDUN Framework}

The LINDDUN framework is a comprehensive tool for conducting privacy threat modeling, allowing organizations to systematically identify and mitigate privacy risks throughout the software development lifecycle. LINDDUN operates in two main phases:

\begin{itemize}
    \item \textbf{Problem Space}: This phase involves creating Data Flow Diagrams (DFDs), mapping privacy threats to elements within the DFD, and identifying potential threat scenarios. DFDs offer a visual representation of how data flows through the system, highlighting points where privacy risks may arise.
    
    \item \textbf{Solution Space}: In this phase, identified threats are prioritized, mitigation strategies are elicited, and corresponding privacy-enhancing technologies (PETs) are selected. LINDDUN's structured methodology ensures that privacy considerations are integrated into both the design and implementation phases of software systems.
\end{itemize}
To serve different needs, LINDDUN comes in various flavors. These approaches vary in complexity and comprehensiveness, ranging from lean to in-depth analysis. LINDDUN framework can be applied through three different methods\footnote{\url{https://linddun.org}}, GO, PRO, and MAESTRO. \textit{LINDDUN GO} takes on a lean, cross-team approach in finding privacy issues. GO comes in the form of a ``card deck" representing the most common privacy threats, with the key hotspots to look for in your system. These self-contained cards will guide you through the privacy assessment. \textit{LINDDUN PRO} takes on a systematic and exhaustive approach in finding privacy issues. The starting point is a DFD system abstraction, where you focus on all interactions between DFD elements and investigate potential privacy threats. Available knowledge support: privacy threat types, privacy threat trees, mapping table.
PRO allows you to leverage tooling to automate your analysis activities. 
Best performed in a structured brainstorming setting with a diverse team of privacy enthusiasts. \textit{LINDDUN MAESTRO} takes on a systematic and exhaustive approach in finding privacy issues by leveraging an enriched system description to enable more precise threat elicitation. Starting point is a threat-specific system abstraction, to support the advanced analysis for threats of that particular type.

By now, only the first two methods of this framework is available. 
Despite its thoroughness, LINDDUN can be resource-intensive, requiring detailed knowledge of system architecture, significant manual effort in diagram creation, and expertise to interpret results effectively. These challenges, while manageable for privacy experts, may present obstacles for organizations lacking specialized resources.

\subsection{Privacy Patterns}

In addition to structured frameworks like LINDDUN, privacy patterns\footnote{\url{https://privacypatterns.org}} have emerged as a useful tool for enhancing privacy practices during system design. Privacy patterns are reusable solutions to common privacy-related challenges, such as data minimization, user consent, and transparency. By adopting these patterns, developers can address specific privacy concerns and improve compliance with privacy regulations. Moreover, the use of privacy patterns encourages a proactive approach to privacy, embedding best practices into system designs from the outset.

\subsection{Large Language Models}

Large Language Models (LLMs) represent a significant advancement in natural language processing and have shown potential for a variety of applications in the field of security. LLMs, such as GPT-4, are built on transformer architectures that allow them to process vast amounts of text data and learn complex linguistic patterns. These models are typically trained on massive datasets, comprising billions of words, enabling them to generate coherent and contextually relevant text. 

LLMs are particularly well-suited for tasks involving the analysis of system documentation, the generation of code or system models, and the identification of security and privacy vulnerabilities. Their ability to understand and generate human-like text makes them a valuable tool for automating processes that traditionally required expert input, such as privacy threat modeling.

However, the use of LLMs is not without challenges. These models can inherit biases from their training data, which raises concerns about fairness and accuracy in their outputs. Additionally, while LLMs excel at generating coherent responses, they may not always produce technically accurate or unbiased results without careful prompt engineering and refinement.

\subsection{Prompt Engineering}

Prompt engineering refers to the design and refinement of input queries to guide the output of LLMs. Well-structured prompts can significantly improve the quality of the responses produced by LLMs, while iterative refinement can be used to further enhance performance.
Key aspects of prompt engineering include:
\begin{itemize}
    \item \textbf{Structure and Clarity}: Clear and structured prompts help guide the model's output towards specific areas of interest, improving the relevance of the generated responses.
    
    \item \textbf{Iterative Refinement}: Prompt engineering often involves multiple iterations to refine the prompt and achieve the desired outcome. This may include adjusting the wording, providing examples, or specifying the format of the output.
    
    \item \textbf{Impact on Performance}: Research has shown that prompt design can significantly affect the performance of LLMs, making prompt engineering an essential skill for developers and researchers working with these models.
\end{itemize}

By leveraging effective prompt engineering, LLMs can be utilized in privacy threat modeling to produce accurate, relevant results with minimal manual effort, reducing the burden on developers and privacy experts while ensuring alignment with ethical and operational standards.

%% file: Sections/3-Tool.tex
\section{Tool Architecture}
\label{sec:tool}

\begin{figure}[]
    \centering
    \includegraphics[width=0.95\linewidth]{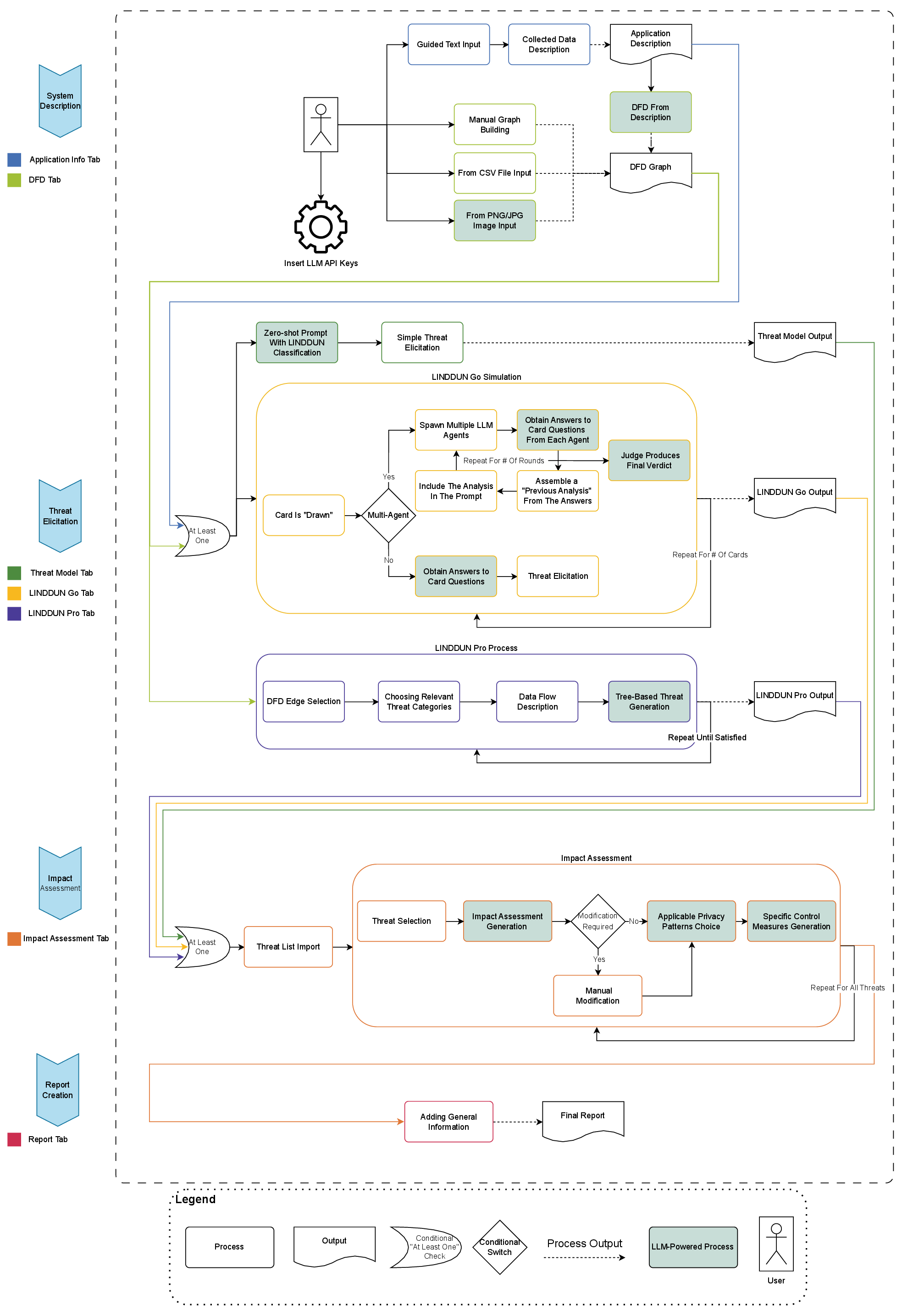}
    \caption{Tool's architecture.}
    \label{fig:tool}
\end{figure}

Figure~\ref{fig:tool} illustrates the tool architecture which has 4 main phases, namely, \textit{System Description}, \textit{Threat Elicitation}, \textit{Impact Assessment}, and \textit{Report Creation}. We used a color code to better specify the processes of each phase. Below we describe each phase in detail.

\subsection{Implementation}

PILLAR\footnote{\url{https://github.com/stfbk/PILLAR}} has been developed with ease of experimentation and functionality in mind. The Python programming language has been devised as the optimal choice for this purpose, due to its high prototyping speed and its widespread adoption as a standard tool for machine learning development and interfacing.
To offer PILLAR users a familiar interface, we developed it as a web application using the Streamlit\footnote{\url{https://streamlit.io}} framework. Many other libraries for the language have been used for specific PILLAR features, such as Graphviz\footnote{\url{https://graphviz.org}} for DFD graph display and modification, \textit{csv} to handle files in the format, \textit{markdown} and \textit{pdfkit} for the final report creation.
Additionally, the LLM functionality is accomplished by using the Python libraries developed by the LLM providers. PILLAR currently supports models offered by OpenAI, Google Gemini and Mistral.  

\subsubsection{Streamlit}
The use of Streamlit proved highly advantageous in creating an interactive interface with minimal coding requirements and allowed to have a clearer outline of the necessary parts of the application. Furthermore, the interface is clear and modern, with intuitive fields and interactivity, as well as fully responsive.
Nonetheless, the framework forces some constraints on the application's appearance and functionality. Some workarounds have been necessary to, for example, defer the report's generation until the download button is pressed, or laying out items correctly in the interface. Of course, these limitations are directly correlated with Streamlit's ease of use and do not significantly compromise the final result.

\subsubsection{LLM providers}
The primary LLM provider used throughout the application is OpenAI\footnote{\url{https://openai.com}}. The company has established itself as the current de facto standard for LLM-based applications and offers robust developer experience for interactions with its models. During PILLAR's development, the \textit{gpt-4o-mini} model has been released, which greatly improves the response quality and cost-effectiveness of our tool compared to its predecessor, \textit{GPT-3.5-turbo}. PILLAR also supports the other OpenAI models, such as \textit{gpt-4o}, for increased accuracy at a higher cost. Another state-of-the-art feature integrated into PILLAR is OpenAI's \textit{Structured Output}: a JSON schema is specified for the LLM output and the correct format is guaranteed to be obtained, all thanks to the recently upgraded API provided by OpenAI.  

For some functionalities, such as the simple threat modeling or multi-agent LINDDUN GO, Google Gemini\footnote{\url{https://gemini.google.com}} and Mistral\footnote{\url{https://mistral.ai}} can be used as LLM providers in PILLAR. These providers offer simple Python integration and similar services to OpenAI, but at the present time do not have comparable structured model output guarantees.




\subsection{Usage}
PILLAR is meant to be operated in a mostly linear fashion, traversing the phases outlined in Figure~\ref{fig:tool}, by a user who has knowledge about the system to be analyzed. An OpenAI API key (with available credit) is required for functionality, while Google Gemini and Mistral keys are only recommended for certain features. All API keys should be inserted in PILLAR's sidebar view. Next, we go through each phase.

\subsubsection{System Description}
The user needs to describe the system to be analyzed, such that the subsequent threat elicitation can be performed. Naturally, the more detailed information provided about the system, including specific technical details and assumptions about the data, the more likely the resulting output will be of higher quality. 

The first way to describe the system's features is by filling out the fields in the \textit{Application Info} tab. These fields guide the user in inserting valuable privacy-related information about the system, useful for the LLM to be more precise in the responses. It is also possible to describe, through a table interface, the types of data collected by the system and other details associated with them, a greatly important aspect of privacy threat modeling.

The system description can also be provided in the form of a Data Flow Diagram (DFD), through the \textit{DFD} tab. The tab allows for manual editing of the DFD through an edge-based table, as well as visualization of the DFD graph. The DFD can also be LLM-generated based on either the textual system description, if already provided, or on an image of the DFD, leveraging \textit{gpt-4o}'s vision capabilities. Once the DFD has been generated from these ways, it can be further perfected and tailored to fit the system.

Both table interfaces allow users to download their contents as CSV files and upload one such file as their new content. This way, they can be edited or generated in another program, or conserved for future analysis on the same system.

These input methods are designed to minimize the effort required for users to describe the system, while still allowing for precision and refinement if desired. One of the biggest challenges in threat modeling is getting developers to engage with and follow the necessary steps in the process, which often requires significant work and security knowledge \cite{dhillon2011developer}. PILLAR aims to alleviate this issue with LLMs' support and ease of use, streamlining the process and reducing the burden on users.

\subsubsection{Threat Elicitation}
After the system description has been provided, threats have to be elicited following the LINDDUN framework.
With either the textual system description or the DFD (or both), a simple threat modeling and LINDDUN GO are available. LINDDUN PRO requires the DFD as an input.

In the \textit{Threat Model} tab, a basic zero-shot privacy threat model can be obtained. The LLM uses the system description provided and identifies threats with a focus on each of LINDDUN's threat categories. The output is generally unspecific but can be a quick initial insight on the types of threats to be aware of, while still minimizing the required effort from the user.

In the \textit{LINDDUN Go} tab, the whole process described by the LINDDUN GO method can be simulated through the use of LLMs. The user can specify the amount of cards to extract from the LINDDUN GO deck and whether or not to carry out a multi-agent simulation. Then, cards up to the number specified are selected from the deck, carrying out the process for each card.

In the single-agent simulation, each card's description and information is provided to the LLM, together with the system description. The LLM's task is to determine whether or not the threat contained in the card is present or not in the system and the reason for either decision. Therefore, the output specifies which threats are relevant and why they should be taken into consideration.

The multi-agent simulation enacts not only GO's card-based elicitation but also the team brainstorming aspect of the methodology. For each card, different LLM agents are spawned, each with a different prompt that suggests its area of expertise and focus, just like real-life members of a privacy threat modeling team. Each of them carries out the single-agent analysis focusing on the aspects important to their own area. 
Once all answers have been collected, they are gathered in a \textit{previous analysis}. Multiple rounds of the analysis are then performed and every time the \textit{previous analysis} obtained from the previous round is supplied as additional input to each agent. 
This process recreates a discussion between the LLMs, providing a chance to share the differing opinions resulting from different aspects of the threat, and increasing the accuracy of the elicitation process. After the specified number of rounds is reached, the last \textit{previous analysis} is supplied to a judge LLM agent, whose task is only to determine the final verdict after taking into account the final opinions from all agents.

During the multi-agent analysis, if specified, LLM agents are randomly selected from the different providers, to add variability in the output and raise more interesting views.

The \textit{LINDDUN PRO} tab makes it possible to perform the full LINDDUN PRO process based on the system DFD. In this method, the user selects an edge to analyze and specifies one or more LINDDUN categories to assess potential threats. 
A description of the specific data flow also needs to be added for the LLM to understand how data is handled throughout the system. 
Once all of the information is provided, the user can obtain a probable threat of the specified category at the different locations, namely source, data flow, and destination. The process can be repeated many times, with different categories and other DFD edges.

Under the hood, PILLAR uses the LINDDUN mapping table\footnote{\url{https://downloads.linddun.org/tutorials/pro/v0/mappingtable.pdf}} to decide whether a threat assessment is applicable for a certain DFD edge.
Furthermore, the LLM receives a version of LINDDUN's threat trees\footnote{\url{https://linddun.org/threat-trees}} and bases its analysis on their nodes.
When presenting the threat, the output also includes the specific threat tree node where the LLM identified the threat.

\subsubsection{Impact Assessment}
Once threats have been elicited, with any one of the methodologies offered by PILLAR, in the \textit{Impact Assessment} tab the user can import the threats found to proceed with the privacy impact assessment. If multiple methodologies were used, the user can select one of them as the source for the threats. 

The threats can be browsed through the interface and it is possible to choose whether or not to include them in the final report. With the LLM, it is possible to generate the impact assessment for the threat, which can also be modified by the user as they see fit. 

Control measures for the threat are generated using privacy patterns. Initially, the LLM receives as input the threat, the system description and a brief description of each privacy pattern. This approach minimizes token usage while allowing the LLM to generate a list of potentially relevant patterns for addressing the threat. Subsequently, a more detailed request is sent to the LLM with the full information for the selected privacy patterns, and the LLM performs a further selection among them, as well as offering reasons for each pattern’s relevance and guidelines on how to implement it within the system.

\subsubsection{Report Creation}
PILLAR offers, as a final result for the user, a PDF report containing all the elicited threats that were chosen in the previous impact assessment, as well as the additional information associated to each of them. Additionally, the report contains some general information, which needs to be provided in the \textit{Report} tab. If it is deemed useful, the report can also include the DFD graph.   

Once it is downloaded, the report can be referenced as either a starting point for further privacy threat modeling or as a guide to contain the elicited privacy threats.

%% file: Sections/5-Relatedwork.tex
\section{Related Work}
\label{sec:related}

The integration of LLMs into the field of cybersecurity has garnered considerable attention due to their ability to process and analyze vast datasets. As cyber threats become more sophisticated, the cybersecurity domain increasingly turns to these advanced models to strengthen defenses. Cybersecurity professionals continually seek innovative solutions to implement robust policies and enhance technological safeguards~\cite{kaur}. These efforts are essential for preventing the unauthorized disclosure of sensitive information, unauthorized access, and various forms of data manipulation~\cite{yang}. 

The ability of LLMs to adapt and scale is particularly valuable for addressing the growing complexity of cyber threats. The following sections highlight key applications of LLMs in cybersecurity, summarize findings from recent research, and discuss notable tools leveraging LLMs to improve security. Applications of LLMs in cybersecurity include:

\begin{itemize}
    \item \textbf{Vulnerability Detection:} LLMs can analyze code and security advisories to identify potential vulnerabilities in software systems.
    \item \textbf{Malware Analysis:} These models assist in classifying and analyzing malware by recognizing complex data patterns.
    \item \textbf{Network Intrusion Detection:} LLMs can process network traffic data to detect anomalies that may signal potential intrusions.
    \item \textbf{Phishing Detection:} By analyzing textual patterns, LLMs can detect phishing attempts with greater precision than traditional methods~\cite{xu2024large, motlagh}.
\end{itemize}

Although the application of LLMs shows promising advancements, research suggests challenges remain, such as the need for larger training datasets and more interpretable models. Future work is expected to focus on improving model explainability, addressing privacy concerns, and developing proactive defense mechanisms against cyber threats.

One significant application of LLMs in network security is web fuzzing. For example, Liang et al.~\cite{liang2023generative} proposed GPTFuzzer, which uses an encoder-decoder architecture to generate effective payloads for web application firewalls (WAFs). It specifically targets vulnerabilities such as SQL injection, cross-site scripting (XSS), and remote code execution (RCE) by generating fuzz test cases. Another notable use of LLMs is in detecting network traffic anomalies. Liu et al.~\cite{liu2023malicious} developed a method to detect malicious URLs by leveraging LLMs to extract hierarchical features. This work extends the use of LLMs in intrusion detection tasks to the user level, demonstrating their generality and effectiveness in intrusion and anomaly detection.

As cyber threats grow in complexity, traditional Cyber Threat Intelligence (CTI) methods struggle to keep pace. AI-based solutions, including LLMs, offer an opportunity to automate and enhance several tasks, ranging from data ingestion to resilience verification. LLMs have shown promise in generating CTI from various sources, including network security texts (e.g., books, blogs, news)~\cite{aghaei2022securebert}, generating structured reports from unstructured data~\cite{siracusano2023time}, and extracting intelligence from network security entity graphs~\cite{perrina2023agir}. Aghaei et al.~\cite{aghaei2023automated} introduced CVEDrill, which generates priority recommendation reports for cybersecurity threats and predicts their potential impact. Furthermore, Moskal et al.~\cite{moskal2023llms} examined the use of ChatGPT in assisting or automating decision-making in response to threat behaviors, illustrating the potential of LLMs in handling simple network attack scenarios.

In the context of vulnerability detection, Liu et al.~\cite{liu2023harnessing} proposed LATTE, which combines LLMs with automated binary taint analysis. LATTE overcomes the limitations of traditional taint analysis, which often requires manual customization of taint propagation and vulnerability inspection rules. Phishing and scam detection is another area where LLMs demonstrate significant utility. Labonne et al.~\cite{labonne2023spam} highlighted the effectiveness of LLMs in spam email detection, showcasing their superiority over traditional machine learning approaches. Additionally, Cambiaso et al.~\cite{cambiaso2023scamming} presented an innovative study suggesting that LLMs can mimic human interactions with scammers in an automated yet meaningless manner, wasting scammers’ time and resources, and thereby reducing the impact of scam emails.

Detection and Intelligence Analysis for New Alerts\footnote{\url{https://github.com/dwillowtree/diana}} (DIANA) is another recently developed tool that automates the creation of detections from threat intelligence using LLMs. In the context of modeling security threats, STRIDE-GPT\footnote{\url{https://github.com/mrwadams/stride-gpt}} is another advanced tool that generates threat models and attack trees for a given application based on the STRIDE methodology. AttackGen\footnote{\url{https://github.com/mrwadams/attackgen}} is a cybersecurity incident response testing tool that leverages the power of large language models and the comprehensive MITRE ATT\&CK framework\footnote{\url{https://attack.mitre.org/}}. The tool generates tailored incident response scenarios based on user-selected threat actor groups and your organisation's details.  

A framework that leverages multiple intelligent LLM agents working collaboratively can enhance the efficiency and effectiveness of handling complex tasks. This approach addresses various challenges such as security risks, scalability, and system evaluation, showcasing the potential for advanced applications in cybersecurity~\cite{talebirad2023multi}. For instance, authors in~\cite{song2024audit} introduced Audit-LLM, a multi-agent log-based insider threat detection (ITD) framework comprising three collaborative agents, where it detects malicious user activities by auditing log entries.

%% file: Sections/6-Conclusion.tex
\label{sec:conclusion}
\section{Conclusion}

In this paper, we presented \textbf{PILLAR}, an AI-powered privacy threat modeling tool designed to automate and enhance the privacy risk identification process by integrating the LINDDUN framework with large language models. PILLAR addresses the limitations of traditional privacy threat modeling approaches, which often require substantial manual effort, expert knowledge, and detailed system descriptions.

By automating key processes such as data flow diagram generation, threat identification, and risk prioritization, PILLAR reduces the workload for developers and privacy experts, while improving the overall efficiency and accuracy of the threat modeling process. The tool’s ability to generate and analyze privacy threats from natural language descriptions of systems demonstrates its potential to streamline privacy assessments, particularly for organizations that may lack specialized privacy resources.

PILLAR introduces innovative features, such as multi-agent collaboration using LLMs, which simulates expert brainstorming sessions to produce more comprehensive and precise privacy threat analyses. Additionally, the incorporation of advanced prioritization techniques ensures that users can focus on addressing the most critical privacy threats, helping to optimize resource allocation.

Moving forward, future enhancements to PILLAR will include refining the accuracy of threat generation and expanding its interoperability with other privacy and security tools. Additionally, we aim to explore deeper integrations of LLMs to further reduce manual input and improve threat modeling accuracy. By addressing these goals, PILLAR will continue to evolve as a robust tool for privacy threat modeling, enabling more organizations to secure their systems and comply with privacy regulations such as GDPR.